\documentclass[11pt]{article}
\usepackage{amsmath}
\usepackage{graphicx,psfrag,epsf}
\usepackage{enumerate}

\usepackage{url} 
\usepackage{bm}
\usepackage{amsthm}
\usepackage{setspace}
\usepackage{color}
\usepackage{times} 
\usepackage{natbib}
\usepackage{amssymb}

\begin{document}


  \title{\bf Assurance for sample size determination in reliability demonstration testing}
  \author{Kevin J Wilson, Malcolm Farrow\\
    School of Mathematics and Statistics, Newcastle University, UK}
\maketitle

\begin{abstract}
Manufacturers are required to demonstrate products meet reliability targets. A typical way to achieve this is with reliability demonstration tests (RDTs), in which a number of products are put on test and the test is passed if a target reliability is achieved. There are various methods for determining the sample size for RDTs, typically based on the power of a hypothesis test following the RDT or risk criteria. Bayesian risk criteria approaches can conflate the choice of sample size and the analysis to be undertaken once the test has been conducted and rely on the specification of somewhat artificial acceptable and rejectable reliability levels. In this paper we offer an alternative approach to sample size determination based on the idea of assurance. This approach chooses the sample size to answer provide a certain probability that the RDT will result in a successful outcome. It separates the design and analysis of the RDT, allowing different priors for each. We develop the assurance approach for sample size calculations in RDTs for binomial and Weibull likelihoods and propose appropriate prior distributions for the design and analysis of the test. In each case, we illustrate the approach with an example based on real data.
\end{abstract}

\noindent%

\newpage

\section{Introduction}

Hardware products are required to have high levels of reliability, particularly those which provide safety-critical functions within larger systems. It is therefore crucial that the users of hardware products have confidence in the reliability claimed by the manufacturers. A typical and widely used approach to provide this confidence is through the use of a reliability demonstration test (RDT), in which a number of the hardware products are put on test and the number which fail or failure times are observed. If the reliability meets some pre-defined threshold then the test is said to be passed and the reliability of the product is demonstrated \citep{Els12}.

An additional consideration affecting the test plan is the type of hardware product of interest. In this paper we classify hardware products into two types: products which are required to function on demand and can either work or fail, such as a parachute, and products which are required to function for a stated period of time, such as an engine. The design of an RDT for failure on demand products is given by the number of items put on test and the number of failures which are allowed for the test to be passed. For time to failure products the RDT design consists of the number of items to put on test and, if an accelerated test is to be used, the stresses of each of the items.  

Calculating sample sizes for RDTs has been considered at least as far back as the 1960s. A traditional approach was to choose the sample size to give a specified power of a hypothesis test (see Section 3 of \cite{Mee04}). Other more recent approaches for more complex tests have been based on properties of hypothesis tests or confidence intervals, e.g., for log-location-scale distributions with failure censoring in \cite{Mck05} and for testing highly reliable products under extreme conditions in \cite{Mea06}.

Another widely investigated approach is based on the idea of risk criteria, which evaluate the risk to the producer and consumer of the product associated with incorrect conclusions from the RDT. The first attempt was the classical risk criteria which defined the producer's (consumer's) risk as the probability of failing (passing) the test conditional on a chosen (un)acceptable value for the reliability (see, for example, Chapter 10 of \cite{Mar82}). To overcome the need to specify a value for the reliability of a passed (failed) test, average risk criteria \citep{Eas70}, which condition on being above (below) a reliability threshold, and Bayesian risk criteria (see, for example, Chapter 10 of \cite{Ham08}), which calculate the probability of (not) reaching a reliability target conditional on passing (failing) the RDT, have been proposed. As a result of the structure of the approach, the prior distribution used in the Bayesian risk criteria is the same for the design and analysis of the RDT.

Assurance (also known as expected power and average power), has been proposed \citep{Spi94,Oha01} as the correct Bayesian method for calculating sample sizes in clinical trials. It is based on the notion that the sample size should be chosen to meet a threshold for the probability that the trial leads to a successful outcome. If the analysis to be conducted following the trial is a frequentist hypothesis test for the superiority of a new treatment over the standard treatment, then the successful outcome would be rejection of the null hypothesis in the test. \cite{Spi94} considered simple structures of clinical trials and \cite{Oha05} extended the idea to trials involving binary data and non-conjugate prior distributions. \cite{Ren13} detailed assurance calculations for clinical trials with time-to-event outcomes, considering exponential and Weibull survival distributions. \cite{Mil18} compared assurance to traditional power calculations in clinical trials for rare diseases.

There is no limitation with assurance that the analysis following the trial be a frequentist test, however. \cite{Oha01} proposed assurance for sample size determination based on a Bayesian analysis following the test. In this case a successful test was defined based on the posterior probability of the new treatment being superior to the standard treatment. They proposed to use different priors in the design and the analysis of the trial, representing the beliefs of different groups of people. \cite{Wal15} use this approach in a case study set in early drug development, and discuss suitable design and analysis priors. \cite{Mui13} also proposed assurance, which they termed ``probability of a successful trial'' for sample size determination based on a Bayesian analysis following the trial. They argued for the same prior distribution to be used in the design and analysis of the trial.  

In this paper we consider the use of assurance for sample size determination outside the context of medicine for the first time. We consider RDTs for products which are required to function on demand and for products which are required to function for a stated length of time. In both cases, we develop an approach based on assurance to select both the number of items to test and the criteria for a successful RDT. We propose suitable structures for the prior distributions for the unknown model parameters and detail how historical data can be incorporated into the assurance calculation.

One important advantage of assurance over typical Bayesian risk-criteria based approaches to RDT design is that we can choose to incorporate the prior judgements of the producer of the hardware product into the design of the RDT, through the design prior, without also imposing them into the analysis of the test results, for which we can use a conservative analysis prior or an analysis prior based on the beliefs of the consumer of the product. We could also, if we wished, base the analysis of the test data on a frequentist hypothesis test, using this to choose the threshold value between test success and failure. Assurance also allows us to incorporate historical data into the design of the RDT, again without imposing these historical data into the analysis. This is in contrast to the risk-criteria based approaches.

Following this introduction, we detail our approach to sample size determination using assurance for failure on demand data, using a binomial likelihood function, in Section \ref{sec:bin}, and for time to failure data, through a Weibull likelihood, in Section \ref{sec:wei}. We then illustrate the approach developed in the context of examples, first for failure on demand data from emergency diesel generators in Section \ref{sec:ex1} and then for time to failure data on pressure vessels in Section \ref{sec:ex2}. The paper concludes with Section \ref{sec:conc}, which provides a summary of the approach and some areas for further work.

\section{Binomial reliability demonstration testing}
\label{sec:bin}

\subsection{Traditional approaches}
\label{trad}

Suppose that we have $n$ items which we are to put on test and that $Y$ is the number which will fail the test. Also define $\pi$ to be the probability that an item survives the test. In reliability demonstration testing, for a given $n$, we define a maximum allowed number of failures $c$. If the actual number of failures in the test exceeds $c$ then the test is failed. If not, the test is passed.

Therefore, a reliability demonstration test plan \citep{Ham08} is given by the pair $(n,c)$. Provided that the failures are independent and identically distributed, the likelihood associated with the test is binomial
\begin{displaymath}
Y\mid\pi\sim\textrm{bin}(n,1-\pi).
\end{displaymath}

Traditionally, a reliability demonstration test (RDT) would be analysed using a hypothesis test where $H_0:\pi=\pi_T$, the quantity $\pi_T$ is the target reliability and $H_1:\pi>\pi_T$. For a $100(1-\alpha)\%$ critical level, we would reject $H_0$ when $\Pr(Y\leq y\mid\pi=\pi_T)\leq\alpha$. That is if $\sum_{j=0}^y{n \choose j}(1-\pi_T)^j\pi_T^{n-j}\leq\alpha$.
We choose $c$ to be the largest value of $y$ satisfying this inequality.

Other approaches have been suggested, often based on the idea of risk criteria. In these approaches, test plans are chosen to keep the probabilities of making the wrong conclusions from the reliability demonstration test small. The two errors considered are named the producer's risk, which is associated with a failed RDT for a product which meets the reliability target, and the consumer's risk, which is associated with a passed RDT for an item which does not meet the reliability target. Classical risk criteria \citep{Tob95}, average risk criteria \citep{Eas70} and (Bayesian) posterior risk criteria (see, for example, Chapter 10 of \cite{Ham08}) have been proposed. The details of the posterior risk criteria for binomial reliability demonstration testing are given in the Supplementary Material. A summary of the producer's risk and consumer's risk for each of the approaches is given in Table \ref{risk}. In the table, $\pi_0$ is defined as the acceptable reliability level and $\pi_1$ is defined as the rejectable reliability level.

\begin{table}[ht]
\centering
\begin{tabular}{|c|c|c|}\hline
Approach & Producer's risk & Consumer's risk \\ \hline
Classical & $\Pr($Test is failed$\mid\pi=\pi_0)$ & $\Pr($Test is passed$\mid\pi=\pi_1)$ \\
Average & $\Pr($Test is failed$\mid\pi\geq\pi_0)$ & $\Pr($Test is passed$\mid\pi\leq\pi_1)$ \\
Posterior & $\Pr(\pi\geq\pi_0\mid$Test is failed$)$ & $\Pr(\pi\leq\pi_1\mid$Test is passed$)$ \\ \hline
\end{tabular}
\caption{The producer's risk and consumer's risk for three risk criteria approaches to choosing a binomial RDT plan.}
\label{risk}
\end{table}

Often in reliability demonstration testing, test plans can involve running large numbers of very reliable items for long periods of time. \cite{Mee04} proposed using past observations $\bm x$ in the analysis to reduce $n$. They named the new tests which incorporated past observations reliability assurance tests.

\subsection{Assurance}
\label{subs:assurance}
If we consider the posterior risk criteria in the Supplementary Material, the approach conflates two distinct processes: the choice of sample size for the test and the analysis to be undertaken once the test has been conducted. A result of this is that the prior used in the analysis stage is the same as that used in the design stage. This may not be appropriate. The specification of $\pi_1$ and $\pi_0$ is a somewhat artificial convenience.

Assurance has been proposed \citep{Spi94,Oha05,Ren13} as the correct Bayesian approach to sample size calculations in medical statistics. 
Assurance chooses a sample size based on the answer to the question ``what is the probability that the reliability demonstration test is going to result in a successful outcome?''. The probability of a successful test is given by
\begin{eqnarray*}
\Pr[\textrm{Successful test}] & = & \int_{0}^{1}\Pr(Y\leq c \mid\pi)p(\pi)d\pi \\
& = & \int_{0}^{1}\sum_{y=0}^{c}{n \choose y}(1-\pi)^y\pi^{n-y}p(\pi)d\pi,
\end{eqnarray*}
where $p(\pi)$ is the prior probability density for $\pi$ and $\Pr(Y\leq c\mid\pi)$ is the cumulative distribution function of $Y$ evaluated at $c$, which in this case is binomial. We would like this probability to be large. We may wish to choose a minimum target, $\gamma$.

Let us suppose that we have data from previous tests of the form $x_i$ for $i=1,\ldots,I$ where $x_i\mid\pi_i\sim\textrm{bin}(n_i,1-\pi_i)$, and the probabilities of items surviving the test in each case can be thought of as coming from the same prior distribution $\pi_i\sim\textrm{beta}(a,b)$, with hyperparameters $a,b$. For example they could be identical components produced in the same factory being tested at various different locations. Then, let us suppose that our probabilitiy of interest also comes from the same prior distribution $\pi\sim\textrm{beta}(a,b)$ and that we define a hyper-prior distribution over $(a,b)$. The probability of a successful test is now
\begin{eqnarray*}
\Pr[\textrm{Successful test}\mid \bm x] & = & \int_{0}^{1}\Pr(Y\leq c \mid\pi)p(\pi\mid \bm x)d\pi \\
& = & \int_{0}^{1}\sum_{y=0}^{c}{n \choose y}(1-\pi)^y\pi^{n-y}p(\pi\mid\bm x)d\pi.
\end{eqnarray*}
We can sample from the posterior distribution $p(\pi\mid\bm x)$ using Markov Chain Monte Carlo (MCMC) in the following way.
\begin{enumerate}
 \item Generate $N$ posterior draws of $a,b$ of the form $a^{(j)},b^{(j)}$ for $j=1,\ldots,N$.
 \item For $j=1,\ldots,N$ draw $\pi^{(1)},\ldots,\pi^{(N)}$ as
 \begin{displaymath}
 \pi^{(j)}\sim\textrm{beta}(a^{(j)},b^{(j)}).
 \end{displaymath}
\end{enumerate}
Using the draws of $\pi$ from the posterior distribution we can evaluate the probability of a successful test, via Monte Carlo integration, as
\begin{displaymath}
\Pr[\textrm{Successful test}\mid \bm x] \approx \dfrac{1}{N}\sum_{j=1}^N\left[\sum_{y=0}^{c}{n \choose y}(1-\pi^{(j)})^y(\pi^{(j)})^{n-y}\right].
\end{displaymath}
Supposing that, for any $n$, we have a way to choose $c$, then we can use assurance in this way to choose the sample size $n$. The critical number of failures $c$ is chosen based on the analysis to be carried out following the test. This flexible approach allows the use of a frequentist hypothesis test or a Bayesian analysis to be used to decide on the success or failure of the test. If a Bayesian approach is to be used to analyse the test data, the prior for the analysis can be different to that used here in the design.

\subsection{Cut-off choice}

\subsubsection{Frequentist approaches}

Consider the binomial test in Section \ref{trad}. We would reject $H_0$ given $Y=y$ if $\sum_{j=0}^{y}{n \choose j}(1-\pi_T)^j(\pi_T)^{n-j}\leq\alpha$, with e.g. $\alpha=0.05$. Therefore we simply select $c$ to be the largest value of $y$ for which this is true.

An alternative to the exact binomial test is to use a normal distribution to approximate the binomial distribution. In this case the test statistic is $Z(y)=[y/n-\pi_T]/[\sqrt{\pi_T (1-\pi_T)/n}]$ and we would choose the largest $c$ such that $Z(c)<Z_{\alpha}$, where $Z_{\alpha}$ is the $100 \alpha$\% critical value of the standard normal distribution.

\subsubsection{Bayesian approaches}

Suppose that, in the analysis of the test result, we have a prior probability density $p_A(\pi)$  for $\pi$. This may be different from the design prior $p(\pi)$. We may choose a rule that, if the posterior probability that $\pi\leq\pi_T$ is small, then the test is passed. That is, the test is passed if $\Pr_A(\pi\leq\pi_T\mid Y=y)\leq0.05$, 
for example, where $\Pr_{A}$ denotes the probability based on the analysis prior.

In this case we would choose $c$ to be the largest value for which $\Pr_A(\pi\leq\pi_T\mid Y=c)\leq0.05$. This posterior probability can be calculated as
\begin{eqnarray*}
\Pr{}_{A}(\pi\leq\pi_T\mid Y=c) & = & \dfrac{\int_{0}^{\pi_T}f(c\mid\pi)p_A(\pi)d\pi}{\int_{0}^{1}f(c\mid\pi)p_A(\pi)d\pi}, \\
& = & \dfrac{\int_{0}^{\pi_T}{n \choose c}(1-\pi)^{c}\pi^{n-c}p_A(\pi)d\pi}{\int_{0}^{1}{n \choose c}(1-\pi)^{c}\pi^{n-c}p_A(\pi)d\pi},
\end{eqnarray*}
where $f(c\mid\pi)$ is the probability mass function of $Y$ evaluated at $c$. For example, if the prior distribution for $\pi$ was chosen to be $\pi\sim\textrm{beta}(\alpha,\beta)$ then the posterior distribution is $\pi\mid Y=c\sim\textrm{beta}(\alpha+n-c,\beta+c)$ and $c$ is chosen using
\begin{displaymath}
\Pr{}_{A}(\pi\leq\pi_T\mid Y=c) = \dfrac{B(\pi_{T};\alpha+n-c,\beta+c)}{B(\alpha+n-c,\beta+c)},
\end{displaymath}
where $B(\cdot;\cdot,\cdot)$ and $B(\cdot,\cdot)$ are the incomplete beta function and beta function respectively.

This approach produces a reliability demonstration test plan, as the data have been used to calculate the sample size $n$ only and not to analyse the results of the test. A reliability assurance test plan can also be produced in this way by utilising the data $\bm X=\bm x$ here in the choice of $c$. This can be achieved using MCMC in the same way as when calculating $n$ from known $c$ in Section \ref{subs:assurance}.

\subsection{Choice of priors}
\label{subs:choicebinom}
In order to use assurance to decide on the optimal test plan $(n,c)$ we need a design prior distribution for $\pi$, $p(\pi)$, to be used in the assurance calculation and, if a Bayesian analysis is to be undertaken following the test, a second prior distribution on $\pi$, $p_A(\pi)$, to be used in the analysis.

The design prior distribution to be used in the assurance calculation, $p(\pi)$, should represent the beliefs of the producer of the items on test. They take the risk associated with the failure of the test and so it is their probability that the test will be a success which will specify the sample size.

The design prior distribution therefore needs to be defined in terms of quantities about which we could reasonably ask an engineer. The beta distribution parameters $(a,b)$ are not suitable. However, we can reparameterise the beta distribution so that $\pi\sim$beta$(mp,m(1-p))$, where $p=a/(a+b)$ is the mean and $m=a+b$ is the size of the ``prior sample'' on which the mean is based. We can then define suitable hyper-prior distributions on $(p,m)$, such as
\begin{eqnarray*}
p & \sim & \textrm{beta}(a_p,b_p), \\
m & \sim & \textrm{gamma}(a_m,b_m).
\end{eqnarray*}

In the analysis following the test, basing the prior on the beliefs of the producer would typically be a controversial choice. There could be several different groups of people who will be be affected by the decisions made following the success or failure of the test. Therefore, it may be more reasonable in designing the test to suppose that a relatively conservative analysis prior will be used. \cite{Spi04} suggest an approach to this: the sceptical prior.

The sceptical prior should be designed so that there is only a small prior probability that $\pi>\pi_T$. So, for example, if a beta$(\alpha,\beta)$ distribution is assumed for $p_A(\pi)$, then we might choose $(\alpha,\beta)$ to satisfy $1-\dfrac{B(\pi_T;\alpha,\beta)}{B(\alpha,\beta)}=\delta$, where $\delta=0.05$.

An alternative to a simple beta distribution which provides more flexibility in the choice of the consumer's prior is to use a
mixture distribution for the analysis prior. This could take the form
\begin{displaymath}
p_{A}(\pi)=\sum_{m=1}^M  q_{m} p_{A,m}(\pi),
\end{displaymath} 
where $p_{A,m}(\pi)$ is the density of mixture component $m$, $q_{m}>0$ for $m=1, \ldots ,M$ and $\sum_{m=1}^M q_{m}=1$. In this case, the posterior probability to assess whether the test is passed becomes
\begin{eqnarray*}
\Pr{}_A(\pi\leq\pi_T\mid Y=c) & = & \dfrac{\int_{0}^{\pi_T}f(c\mid\pi)\sum_{m=1}^M q_{m} p_{A,m}(\pi)\ d\pi}{\int_{0}^{1}f(c\mid\pi)\sum_{m=1}^M q_{m} p_{A,m}(\pi)\ d\pi}, \\
& = & \dfrac{\sum_{m=1}^M \left[q_{m}\int_{0}^{\pi_T}{n \choose c}(1-\pi)^{c}\pi^{n-c}p_{A,m}(\pi)\ d\pi\right]}{\sum_{m=1}^M\left[q_{m}\int_{0}^{1}{n \choose c}(1-\pi)^{c}\pi^{n-c}p_{A,m}(\pi)\ d\pi\right]},
\end{eqnarray*}
and if each of the component distributions in the mixture is a beta distribution, then this can be expressed in terms of beta functions as previously. 

Assuming that the producer does not know what the consumer's prior is, another way to think of this is that the producer has a joint prior over $\pi$ and the consumer's prior, or, at least, the parameters of the consumer's prior. For example, the producer might think that the consumer's prior density is one of $ p_{A,1}(\pi), \ldots, p_{A,M}(\pi)$ and assign probability $q_{m}$ to $p_{A,m}(\pi)$. 

Given that the consumer actually has prior $m$, this leads to a cut-off $c_{m}$, exactly as for a single-component prior. So we now have 

\[ \Pr[\textrm{Successful test}\mid \bm x] = \sum_{m=1}^{M}\left[ q_{m} \int_{0}^{1} \Pr(Y \leq c_{m} \mid \pi) p(\pi \mid \bm x)\ d \pi \right]. \]

In effect, we now have a producer's probability distribution for the cut-off $c$, with $ \Pr(c=j) = u_{j}$, $J_{L} \leq j \leq J_{U}$, where $J_{L}$ and $J_{U}$ are the minimum and maximum values taken by $c$ and 
$ u_{j} = \sum_{m : c_{m}=j} q_{m}. $

So

\[ \Pr[\textrm{Successful test}\mid \bm x] = \sum_{j=J_{L}}^{J_{U}} \left[ u_{j} \int_{0}^{1} \Pr(Y \leq j \mid \pi) p(\pi \mid \bm x)\ dx \right]. \]

Let $ \int_{0}^{1} \Pr(Y=j \mid \pi) p(\pi \mid \bm x)\ dx = s_{j} $
and $ U_{j} = \sum_{k=J_{L}}^{j} u_{k} . $
Then $\Pr[\textrm{Successful test}\mid \bm x] = \sum_{j=0}^{J_{U}} U_{j} s_{j}$.

\section{Weibull reliability demonstration testing}
\label{sec:wei}

\subsection{Assurance}

In binomial demonstration testing, we are interested in items which either work or fail on demand. For items which are expected to function continuously for long periods of time Weibull demonstration testing is more appropriate. In this case, we would put $n$ items on test, and record their failure times $t_i$, for $i=1,\ldots,n$. We assume that the time to failure of items follows a Weibull distribution
$T \mid \rho,\beta \sim \textrm{Weibull}(\rho,\beta)$ with shape parameter $\beta$ and scale parameter $\rho$. The probability density function of the Weibull distribution is $f(t\mid \rho,\beta)=\rho\beta (\rho t)^{\beta-1}\exp\left[-(\rho t)^\beta\right]$.

To decide on a reliability demonstration test plan, we need to decide on a metric to assess the reliability of the product. Consider the failure time distribution $F(\cdot)$, and suppose that we were interested in a time $\tau_q$ such that $F(\tau_q)=1-q$. Then $\tau_q$ is known as the reliable life of the product for $q$, i.e. the time beyond which the proportion $q$ of the items survive. In a RDT for time to failure data, we can specify some target $\tau_{q,*}$ for the reliable life at $q$. The test plan is the number of items to put on test, $n$, which we will test until failure. 

However, for highly reliable items, testing until failure in normal operating conditions may not be feasible. In this case, accelerated testing can be done, in which items are tested at a much greater stress (e.g. temperature, pressure, vibration) than that of typical use. Suppose we are to conduct the test at stresses $\bm s_{\rm{test}}=(s_{\rm{test},1},\ldots,s_{\rm{test},n})$ and that our target reliable life is specified under a (typically lower) stress $s_{*}$. We can relate the failure times under different stresses using a link function,
\begin{displaymath}
\log(\rho) = g(s,\bm\theta),
\end{displaymath}
where $g(\cdot,\cdot)$ is the link function representing this relationship and the elements of $\bm\theta$ are the parameters of the relationship.

We can express the reliable life in terms of the parameters of the Weibull distribution. Doing so gives
\begin{displaymath}
\tau_q = \rho^{-1}\left[-\log(q)\right]^{\frac{1}{\beta}}.
\end{displaymath}
The assurance is
\begin{eqnarray*}
\Pr(\textrm{Successful test}) & = & \int_{0}^{\infty}\int_{0}^{\infty}\Pr(\textrm{Test passed}\mid\rho,\beta)p(\rho,\beta)\ d\rho\ d\beta,
\end{eqnarray*}
where $p(\rho,\beta)$ is a joint design prior density for $\rho,\beta$.

We could analyse the test results using a hypothesis test in a similar way to the binomial case. A suitable test in this context would be a maximum likelihood ratio test \citep{Col94} with the null hypothesis being $ \tau_{q}=\tau_{q,*}$ and the alternative being $\tau_{q} > \tau_{q,*}$. 

However, we will suppose that we are to perform a Bayesian analysis following the RDT. One criterion which we could use to declare the test passed is if the analysis posterior probability that $\tau_q\geq \tau_{q,*}$ under stress $s_{*}$, given the observation of failure times $\bm t=(t_1,\ldots,t_n)$ in the test, 
under stresses $\bm s_{\rm{test}}$, is large. 
That is, the test is passed if $\Pr_{A}(\tau_q\geq \tau_{q,*})\geq 1-\delta = 0.95$, for example. To evaluate this quantity we could use MCMC. Note that to make inference about $\bm\theta$, we need to use more than one stress. The likelihood function is $L(\bm t\mid\rho,\beta)=\prod_{i=1}^nf(t_i\mid\rho,\beta)$. For right censored items, which have not failed by the end of the test, the density $f(t_i\mid\rho,\beta)$ is replaced with the reliability $R(t_i\mid\rho,\beta)$ in the likelihood.  

Suppose that the analysis prior distribution is given by $p_A(\beta,\bm\theta)$ and that we have observed failure times $\bm t$ under stresses $\bm s_{\rm{test}}$. Then we could assess the RDT criterion above as follows.
\begin{enumerate}
 \item Sample $\beta_{(i)},\bm\theta_{(i)}$, for $i=1,\ldots,N_{1}$, from their analysis posterior distribution using MCMC.
 \item Evaluate
 \begin{displaymath}
 \rho_{(i)}=\exp\left\{g(s_{*},\bm\theta)\right\}
 \end{displaymath}
 and then find
 \begin{displaymath}
 \tau_q^{(i)} = \rho_{(i)}^{-1}\left[-\log(q)\right]^{1/\beta_{(i)}}.
 \end{displaymath}
 \item Estimate the posterior probability as $\Pr_{A}(\tau_q\geq \tau_{q,*})\approx\dfrac{1}{N_1}\sum_{i=1}^{N_1}I\left[\tau_q^{(i)}\geq \tau_{q,*}\right]$, where $I\left[\tau_q^{(i)}\geq \tau_{q,*}\right]$ is an indicator function which takes the value 1 if $\tau_q^{(i)}\geq \tau_{q,*}$ and 0 otherwise.
 \item Assess whether $r_{q}=\Pr(\tau_q\geq \tau_{q,*})\geq 1-\delta$.
\end{enumerate}
A na\"{i}ve approach to assess the assurance would be to sample $M$ sets of parameters $(\rho^{(k)},\beta^{(k)})$, $k=1,\ldots,M$ from the design prior distribution and, for each, sample $N_2$ sets of hypothetical data $\bm t^{(j,k)}$, $j=1,\ldots,N_2$ from the likelihood $L(\bm t\mid\rho^{(k)},\beta^{(k)})$. The assurance would then be given by the Monte Carlo approximation
\begin{eqnarray}\label{success}
\Pr(\textrm{Successful test}) & \approx & \dfrac{1}{N_2\times M}\sum_{k=1}^M\sum_{j=1}^{N_2}I(r_q\geq0.95\mid\bm t^{(j,k)}),
\end{eqnarray}
where $I(r_q\geq 1-\delta\mid\bm t^{(j,k)})$ is an indicator variable which takes the value 1 if $r_q\geq 1-\delta$ and 0 otherwise. This would involve $M\times N_1\times N_2\times n_{\max}$ calculations to evaluate the assurance for all sample sizes in the range $n=1,\ldots,n_{\max}$. The sample size $n$ would be chosen to be the smallest value which meets a specified level of assurance.

A more efficient way to assess the assurance is to adapt the numerical scheme of \cite{Mul99,Mul96}. In this case we would proceed as follows:

\begin{enumerate}
    \item Select a number $M$ of sample sizes $n_j\in[1,\ldots,n_{\max}]$, for $j=1,\ldots,M$.
    \item Simulate 
    \begin{eqnarray*}
    \left(\rho^{(j)},\beta^{(j)}\right) & \sim & p(\rho,\beta), \\
    \bm t_{n_j}^{(j)} & \sim & L(\bm t\mid\rho^{(j)},\beta^{(j)}),
    \end{eqnarray*}
    where $p(\rho,\beta)$ is the design prior, $L(\bm t\mid\rho,\beta)$ is the Weibull likelihood and $\bm t_{n_i}$ is a vector of failure times of length $n_j$.
    \item Evaluate the vector $\bm x$, where
    \begin{displaymath}
    x^{(j)} = I\left[\Pr{}_A(\tau_q\geq\tau_{q,*}\mid\bm t_{n_j}^{(j)})\geq 1-\delta\right],
    \end{displaymath}
    where $\Pr_A(\tau_q\geq\tau_{q,*}\mid\bm t_{n_j}^{(j)})$ is the posterior probability that $\tau_q\geq\tau_{q,*}$ based on the analysis prior. Note that this represents slightly different notation to the previous page.
    \item We repeat Steps 2 and 3 a small number of times to obtain a proportion $\hat{p}^{(j)}$ of successes for sample size $n_j$.
    \item A smoother is used to fit a curve to $\hat{p}^{(1)},\ldots,\hat{p}^{(M)}$. This is an estimate of the assurance.
\end{enumerate}
This greatly reduces the number of computations required to evaluate the assurance. An alternative would be to use an augmented MCMC scheme \citep{Mul99,Coo08}. This is an area for future work.

We are able to incorporate past observations into the assurance calculation. Suppose we have historical observations $(\tilde{t}_{i,j},\tilde{s}_{i,j})$, which represent times to failure and stress levels at locations $i=1,\ldots,m$ of items $j=1,\ldots,n_i$, and these observations follow Weibull distributions $\tilde{t}_{i,j}\mid\rho_{i,j},\beta\sim\textrm{Weibull}(\rho_{i,j},\beta)$ with scale parameter $\log(\rho_{i,j})=g(s_{i,j},\bm\theta_{i,j})$ and a common shape parameter $\beta$. We can use a hierarchical structure to learn about the parameters of the current test by giving $\bm\theta_{i,j}$ prior distributions $p(\bm\theta_{i,j})$ with common hyper-parameters $\bm\phi_{\bm\theta}$ having prior distribution $p(\bm\phi_{\bm\theta})$ and specifying a prior distribution on $\beta$, $p(\beta)$. We further suppose that the current test has the same prior structure with common $p(\beta), p(\bm\phi_{\bm\theta})$.

We incorporate the historical information into the assurance calculation via
\begin{displaymath}
\Pr(\textrm{Successful test}\mid\tilde{\bm t}) = \int_{0}^{\infty}\int_{0}^{\infty}\Pr(\textrm{Test passed}\mid\rho,\beta)p(\rho,\beta\mid\tilde{\bm t})\ d\rho\ d\beta,
\end{displaymath} 
where $p(\rho,\beta\mid\tilde{\bm t})$ is the joint design posterior distribution for $\rho$ and $\beta$ given the obervations $\tilde{t_{i,j}}$.

We evaluate this assurance in the same way as in the case with no previous observations, with the only change being that we simulate $\beta^{(j)},\bm\theta^{(j)}$ values from the design posterior distribution using MCMC rather than the design prior distribution in Step 2 of the RDT assessment algorithm above. 

\subsection{Choice of priors}
\label{sec:priors}

In order to detail the prior elicitation and specification, we need to specify the function $g(\cdot,\cdot)$. We choose to use the general class 
$$ g(s,\bm\theta) = \alpha_0 + \alpha_1 s^k + \epsilon, $$
where $\alpha_0, \alpha_1$ are intercept and slope terms and $\epsilon$ is a location-specific random effect. Well-known cases of this class are $k=1$ which gives a linear relationship and $k=-1$ which gives an Arrhenius relationship.

For the design prior we wish to elicit the beliefs of engineers about the reliability of the items on test. To do so, we need to ask them questions about observable quantities. As a result of the more complex model structure in the Weibull case than the binomial case, we split the elicitation and specification task into three stages, each of which is outlined below.

However, before that, we outline a suitable structure for the prior. Consider the link function above,
$$ \log\rho_{i,j} = \alpha_0 + \alpha_1s_{i,j}^k + \epsilon_{i}. $$
We choose to give the regression parameters $(\alpha_0,\alpha_1)$ a bivariate Normal prior distribution
$\bm\alpha = (\alpha_0,\alpha_1)^{T}\sim \textrm{BVN}(\bm\mu,\Sigma)$,
for hyper-mean vector $\bm\mu=(\mu_0,\mu_1)^{T}$ and variance matrix $\Sigma$ with diagonal elements $(\sigma^2_{00},\sigma^2_{11})$ and off-diagonal element $\sigma^2_{01}$. 

We give the location-specific effects zero-mean normal prior distributions with a common variance, 
$\epsilon_i \sim \textrm{N}(0,v_{\epsilon})$,
where $v_\epsilon$ is a hyper-parameter to be chosen. To complete the prior specification, we need a prior for the Weibull shape parameter, $\beta$. A gamma distribution is a suitable choice and so we set $\beta\sim$gamma$(a_{\beta},b_{\beta})$.

The result is that there are 8 hyper-parameters to be specified in the design prior. We choose to do this using questions about quantiles of the lifetime distribution in terms of a hypothetical large future sample so that the empirical quantiles are, in principal, observable and, since the sample is large, aleatory uncertainty is dominated by epistemic uncertainty. We rescale the stress values so that $s=0$ is a plausible value.

\begin{description}
\item[Stage 1] We ask the expert to suppose that stress is at a specified level $s$. We ask the expert for their lower quartile, median and upper quartile for the reliable life for two different values of $q$. For example, we ask them about the times by which 1/3 and 2/3 of items will have failed. These are:
\[ \tau_{q}  =  \exp\left\{-(\alpha_0+\alpha_{1}s+\epsilon_i)\right\}\left[-\log(q)\right]^{1/\beta}, \]
for $q=1/3, 2/3$.
Specifically, we ask the expert to make judgements about the ratio
\[ \frac{\tau_{2/3}}{\tau_{1/3}} = \left[ \frac{\log(2/3)}{\log(1/3)} \right]^{1/\beta}.\]
A value for this ratio gives a value for $\beta$:
\[ \beta = \frac{ \log[\log(2/3)/\log(1/3)] }{\log[ \tau_{2/3}/\tau_{1/3}]}. \]

By eliciting the expert's quartiles for the ratio, we obtain quartiles for $\beta$ and we can then choose $ a_{\beta}$ and  $ b_{\beta} $ to match these.

\item[Stage 2] We now ask the expert to consider two different locations, $i,k$ and the reliable life at these locations for the same stress and the same value of $q$, e.g. $q=1/2$. Then, if $\tau_{q,i}(s)$ is the reliable life with probability $q$ at location $i$ and stress $s$, we have
\[ \frac{\tau_{q,i}(s)}{\tau_{q,k}(s)} = \exp \{ \epsilon_{i}-\epsilon_{k} \}. \]
The expert's quartiles for this ratio lead to quartiles for $\epsilon_{i}-\epsilon_{k}$, which has variance $2 v_{\epsilon}$, and hence to a value for $v_{\epsilon}$.

\item[Stage 3] We have
\[ \log[\tau_{q,i}(s)] = -(\alpha_{0} + \alpha_{1}(s) + \epsilon_{i}) + \beta^{-1} \log[-\log(q)]. \]
If we choose $q=e^{-1}$ then the second term on the right vanishes. In elicitation, the value $q=1/3$ is more practical and gives a reasonable approximation since $|\log[-\log(1/3)]|<0.1$, unless $\beta$ is very small. Let $M_{p}[\tau_{q,i}(s)]$ denote the expert's $p$-quantile for $\tau_{q,i}(s)$.Then $\log\{M_{1/2}[\tau_{1/3,i}(0)]\}$ gives $\mu_{0}$ and $\log\{M_{1/2}[\tau_{1/3,i}(0)]\}$ for $s \neq 0$ gives $\mu_{0}+\mu_{1}s$ so we obtain $\mu_{1}$. For the variances we can use, for example, the expert's quartiles. So $\log\{M_{1/4}[\tau_{1/3,i}(0)]\}$ and $\log\{M_{3/4}[\tau_{1/3,i}(0)]\}$ lead to $ \sigma^{2}_{00} + v_{\epsilon}$ and hence to $ \sigma^{2}_{00}$ and $\log\{M_{1/4}[\tau_{1/3,i}(s_{1})]\}$, \\ $\log\{M_{3/4}[\tau_{1/3,i}(s_{1})]\}$, $\log\{M_{1/4}[\tau_{1/3,i}(s_{2})]\}$ and $\log\{M_{3/4}[\tau_{1/3,i}(s_{2})]\}$ lead to two simultaneous equations from which we can find $\sigma^{2}_{11} $ and $\sigma^{2}_{01}$.
\end{description}

In the case where we are to observe historical lifetime data in other locations which can be combined with the design prior distribution to form a design posterior distribution, we may wish to learn about the value of $v_{\epsilon}$. In this case we give $v_{\epsilon}$ an inverse gamma prior distribution so
$v_{\epsilon}^{-1}  \sim  \textrm{gamma}(a_{\epsilon},b_{\epsilon})$,
where  $(a_{\epsilon},b_{\epsilon})$ are hyper-parameters to be chosen. We modify Stage 2. The expert's predictive distribution for  $\tilde{T}=(\epsilon_{i}-\epsilon_{j})/\sqrt{2b_{\epsilon}/a_{\epsilon}}$ is a Student's $t$-distribution on $ 2 a_{\epsilon}$ degrees of freedom. To find both $a_{\epsilon}$ and $b_{\epsilon}$, we need two quantiles corresponding to probabilities $q_{1}$ and $q_{2}$ with $q_{1} \neq q_{2}$, $q_{1} \neq 1- q_{2}$ and $q_{j} \neq 1/2$ for $j=1,2$, e.g. $ q_{1}=0.6$ and $q_{2}=0.8$. Alternatively the expert might imagine a large number of such ratios $\tau_{q,i}(s)/\tau_{q,j}(s)$ and give quartiles for the empirical upper quartile of the ratios.

For the analysis prior, we can assume the same prior structure. A sceptical analysis prior would choose the hyper-parameters to give small prior probability to the event $\tau_{q}>\tau_{q,*}$ under stress $s_{*}$. As in Section \ref{subs:choicebinom}, another possibility is for the producer to assign probabilities to possible consumer's prior distributions.

\section{Examples}

\subsection{Example 1: a binomial reliability demonstration test}
\label{sec:ex1}

We consider an example from \cite{Mar96}. We would like to demonstrate the reliability of an emergency diesel generator in a nuclear power plant, with a target reliability of $ \pi_{T}=0.96$. We will compare assurance based on three analysis methods: the exact binomial test, the Bayesian approach using a simple sceptical analysis prior and the Bayesian approach using a mixture analysis prior. In the binomial test we reject $H_0$ if $p<0.05$ and in the Bayesian approaches we conclude that the test is passed if $\Pr_A(\pi\leq\pi_T\mid Y=y)\leq 0.05$.

\begin{figure}[h!]
\centering
\includegraphics[height=3in]{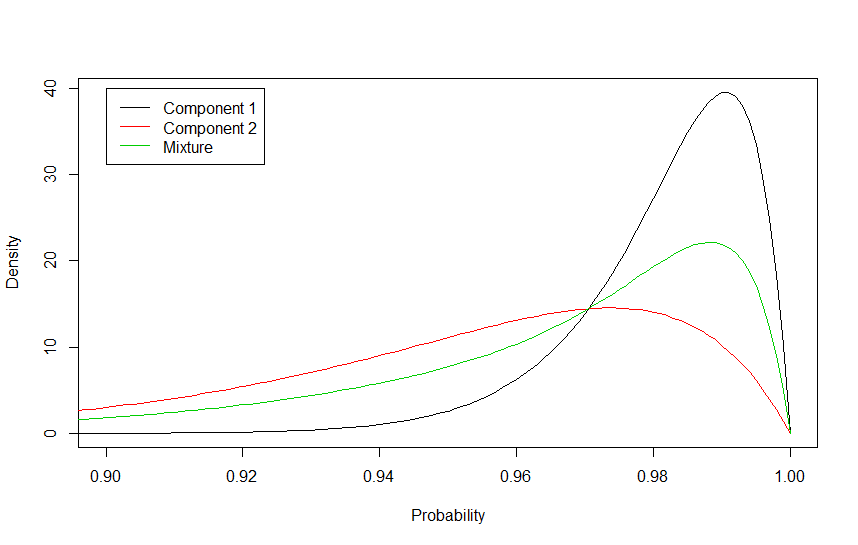}
\caption{The two mixture component priors and the mixture prior for the probability of a generator working on demand.}
\label{mix}
\end{figure}

We choose a beta distribution for the sceptical analysis prior with $\alpha=6.45$ and $\beta=2$. This gives $\Pr_A(\pi>\pi_T) = 0.05$. 
We consider a two-component mixture prior for illustration. The first component gives a probability of 0.75 to the generator meeting the reliability target and the second component gives this event a probability of 0.25. Suitable prior distributions to meet these specifications are beta distributions with parameters $a_1=106, b_1=2$ and $a_2=38, b_2=2$ respectively. We give 60\% weight to mixture component 1 and 40\% weight to mixture component 2. Then
\begin{displaymath}
p_A(\pi)=0.6\dfrac{\Gamma(a_1+b_1)}{\Gamma(a_1)\Gamma(b_1)}\pi^{a_1-1}(1-\pi)^{b_1-1}+0.4\dfrac{\Gamma(a_2+b_2)}{\Gamma(a_2)\Gamma(b_2)}\pi^{a_2-1}(1-\pi)^{b_2-1}.
\end{displaymath}
The two component distributions and the mixture prior are given in Figure \ref{mix}.

The posterior distribution for the mixture given $y$ successes in $n$ trials is of the same form as the prior with the prior weights $p_{1}^{(0)}$ and $p_{2}^{(0)}$ updated to
\begin{displaymath}
p_{1}^{(1)}=\dfrac{\tilde{p}_1}{\tilde{p}_1+\tilde{p}_2} \hspace{1cm} {\rm and} \hspace{1cm}  p_{2}^{(1)}=\dfrac{\tilde{p}_2}{\tilde{p}_1+\tilde{p}_2},
\end{displaymath}
where
\begin{displaymath}
\tilde{p}_i= p_{i}^{(0)}\dfrac{\Gamma(a_i+b_i)}{\Gamma(a_i)\Gamma(b_i)}\dfrac{\Gamma(a_i+y)\Gamma(b_i+n-y)}{\Gamma(a_i+b_i+n)},
\end{displaymath}
and the updated beta distribution parameters are $A_i=a_i+y$ and $B_i=b_i+n-y$ for $i=1,2$.

In the design prior $p(\pi)$ we give $(m,p)$ gamma$(200,1)$ and beta$(78,2)$ priors respectively. This gives a prior mean for $p$ of 0.975 and a prior mean of 200 for the prior sample size. 

\begin{figure}[h!]
\centering
\includegraphics[width=0.49\linewidth]{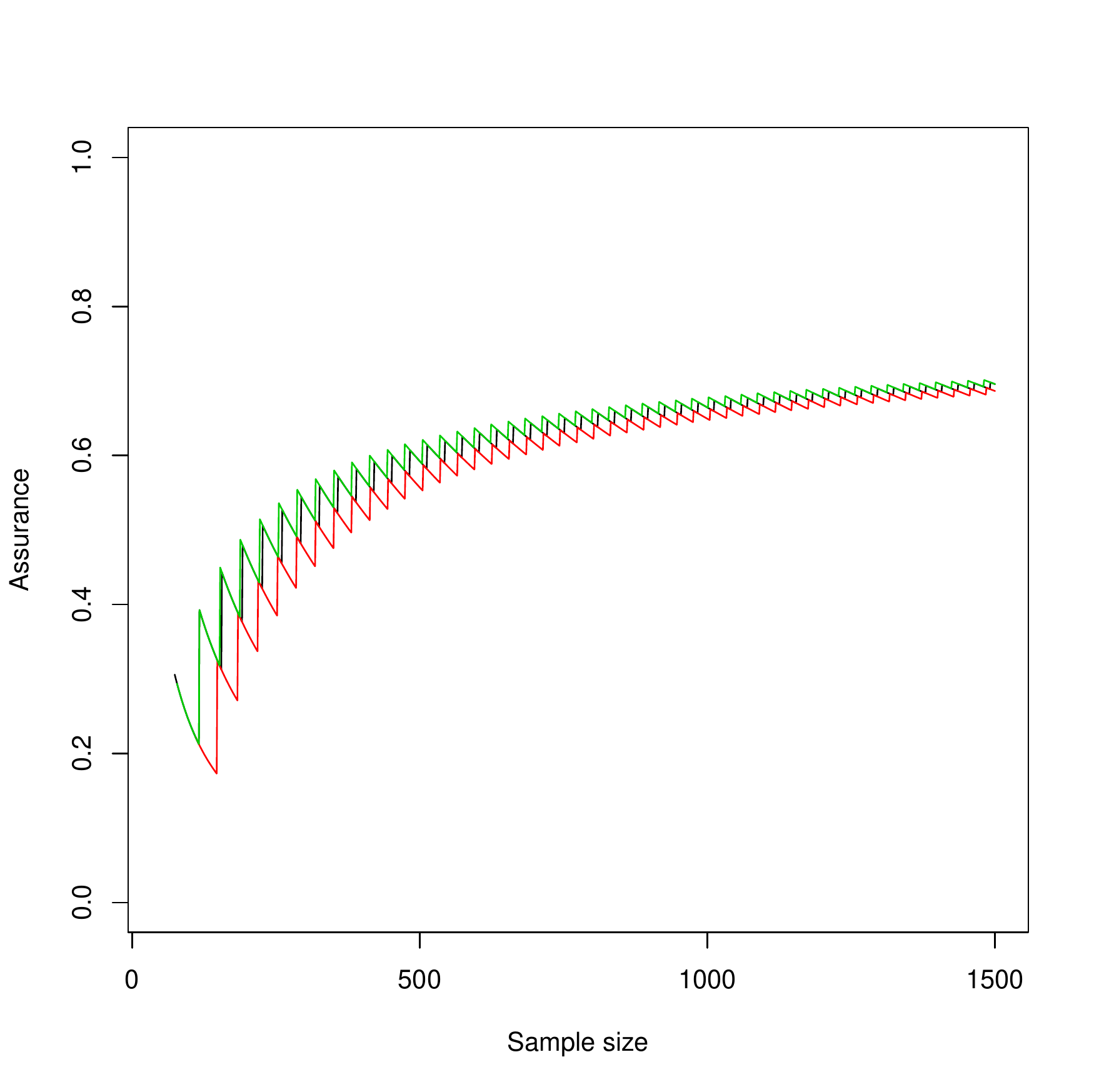}
\includegraphics[width=0.49\linewidth]{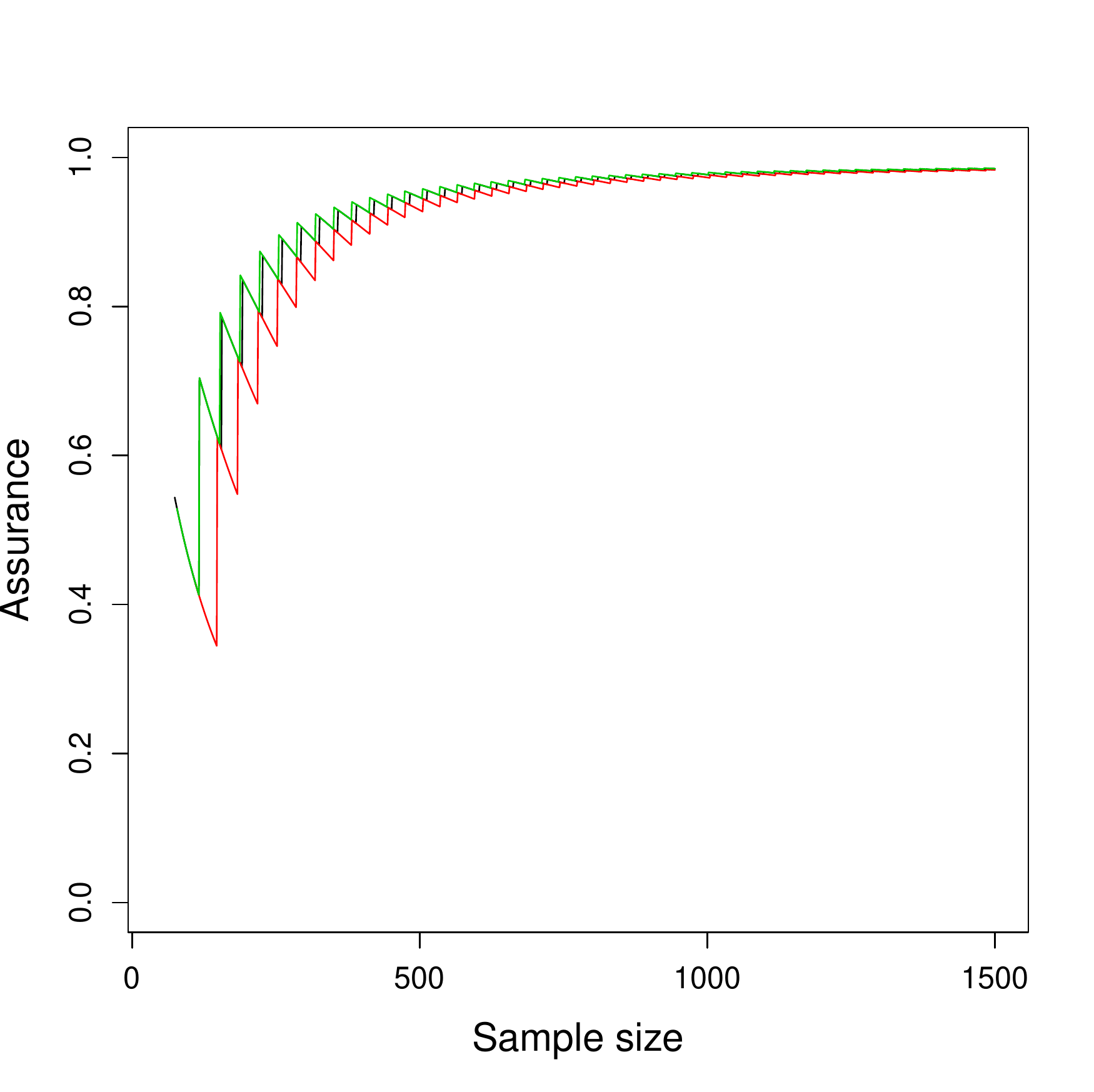}
\caption{The sample size $n$ against the assurance for the binomial test (black), the sceptical prior (red) and the mixture prior (green) based on the design prior distribution (left) and design posterior distribution (right).}
\label{assplot}
\end{figure}

Based on these specifications, we can plot the sample size $n$ against the assurance for each of the three approaches. This is given in the left-hand side of Figure \ref{assplot}.

The binomial test is given in black, the sceptical prior in red and the mixture prior in green. We see that the mixture prior gives the highest assurance for any particular sample size, the binomial test gives the next highest assurance and the sceptical prior the lowest assurance in this case. The lines are decreasing as $c$ remains constant but $n$ increases, and then jump each time $c$ increases. To achieve an assurance of 50\% in this case would require $n=(227,279,222)$ under the three respective methods. We cannot achieve an assurance of greater than 80\% in this case, as this is the prior probability that the target reliability will be met under the design prior. Note that the assurance is slowly converging towards 80\%. For example, with a sample size of $n=10,000$ we achieve an assurance of 76.8\% and with a sample size of $n=100,000$ we achieve an assurance of 79.4\%.

\begin{figure}[h!]
\centering
\includegraphics[height=3in]{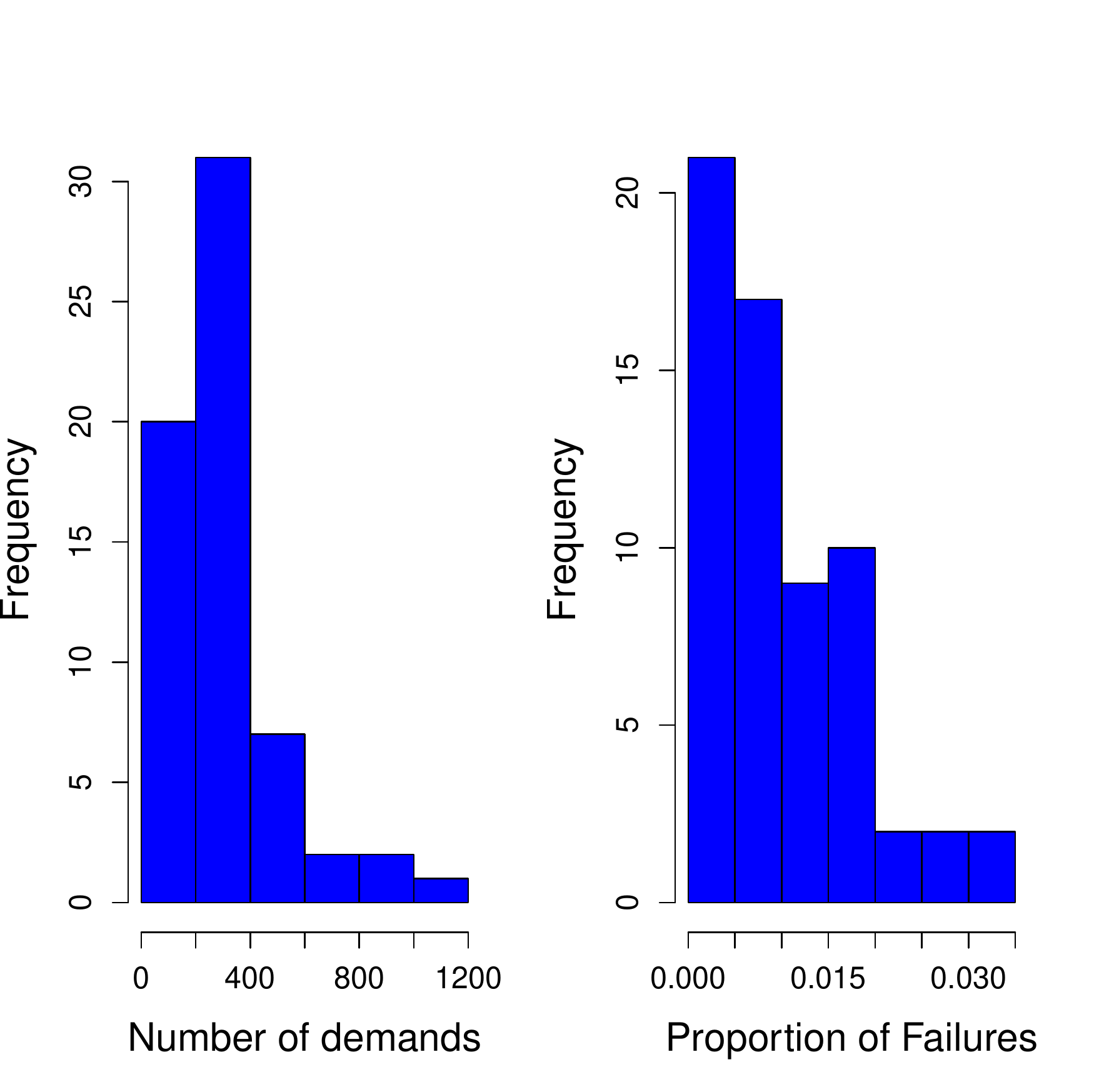}
\caption{The number of demands and proportion of failures of 63 emergency generators in nuclear power plants.}
\label{dem}
\end{figure}

However, we have data on the behaviour on demand of generators of the same type at other plants. The data represent tests of 63 generators in nuclear power plants in the USA. In each case the number of failures on demand of the generator out of the total number of demands was recorded. The number of demands and the proportion of failures in the tests are given in Figure \ref{dem}.
We see that we have some fairly large test data sets and that the failure proportions recorded in the tests are very low, typically below 3\%.

\begin{figure}[h!]
\centering
\includegraphics[height=3in]{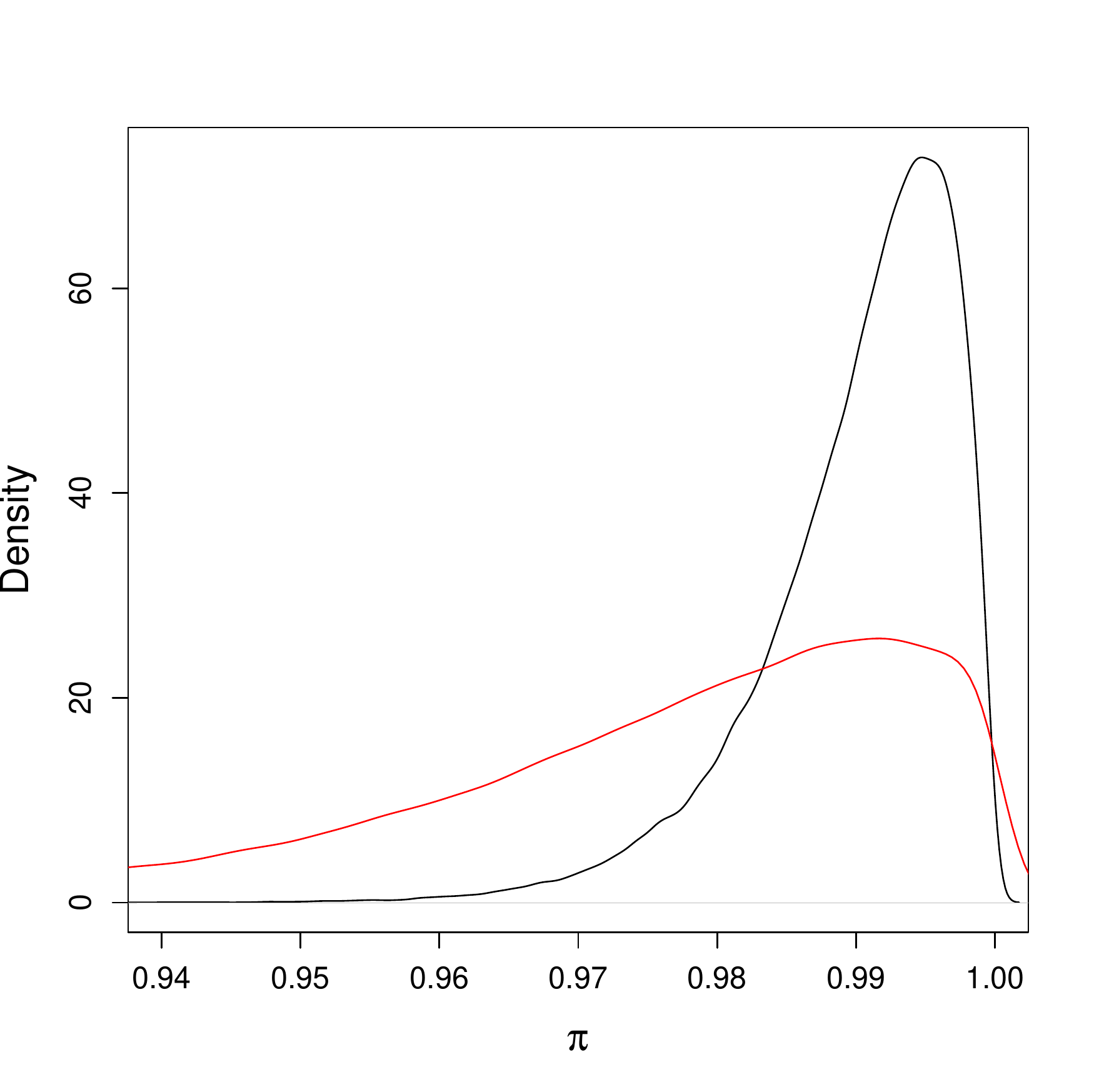}
\caption{The prior distribution (red) and the posterior distribution (black) for $\pi$, the probability of success of the an emergency generator on demand.}
\label{post}
\end{figure}

Using the same design prior as above, we generate 10,000 samples of $(p,m)$ from the posterior distribution using rjags \citep{rja16}, having discarded 1000 samples as burn in. The design posterior distribution of $\pi$, the probability that an emergency generator will work on demand, is given in Figure \ref{post} (in black) alongside the design prior distribution for $\pi$ (in red).

We see that the posterior distribution of the probability of success of a generator is tightly concentrated around very high probabilities, with posterior mean and median of 0.990 and 0.992 respectively. The prior distribution is more diffuse, although much of the density is still concentrated around high probabilities.

We are able to calculate the assurance for each of the three methods as above using this posterior distribution for $\pi$ in place of the design prior. The results for different values of the sample size are given in the right-hand side of Figure \ref{assplot}.

We see, by incorporating the data into the analysis, that we are able to provide a reliability demonstration test which gives assurances of up to almost 100\%. To give assurance of 50\% now only requires sample sizes of $n=(74,141,78)$ for the three analysis methods, all of which are much reduced. Using the sample sizes from the prior calculations would give assurances of $(0.866,0.885,0.872)$ respectively.



\subsection{Example 2: a Weibull reliability demonstration test}
\label{sec:ex2}

\begin{figure}[ht]
\centering
\includegraphics[width=0.45\linewidth]{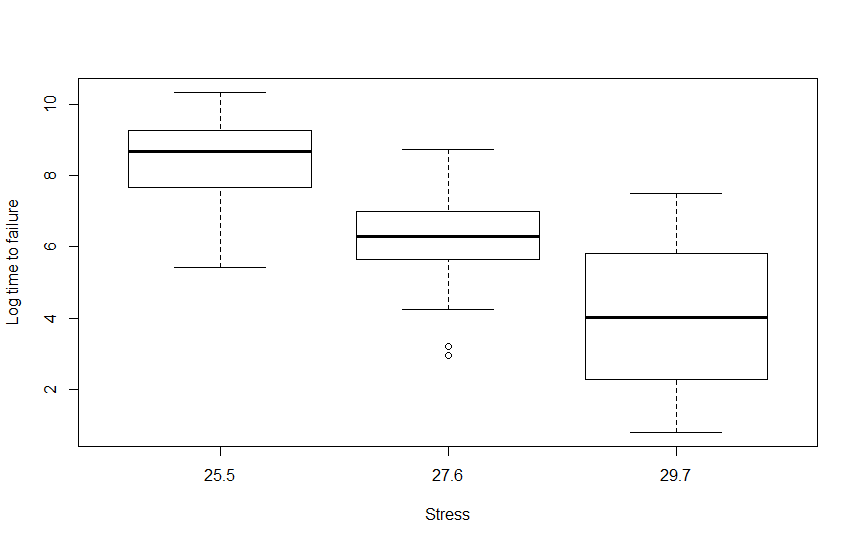}
\includegraphics[width=0.45\linewidth]{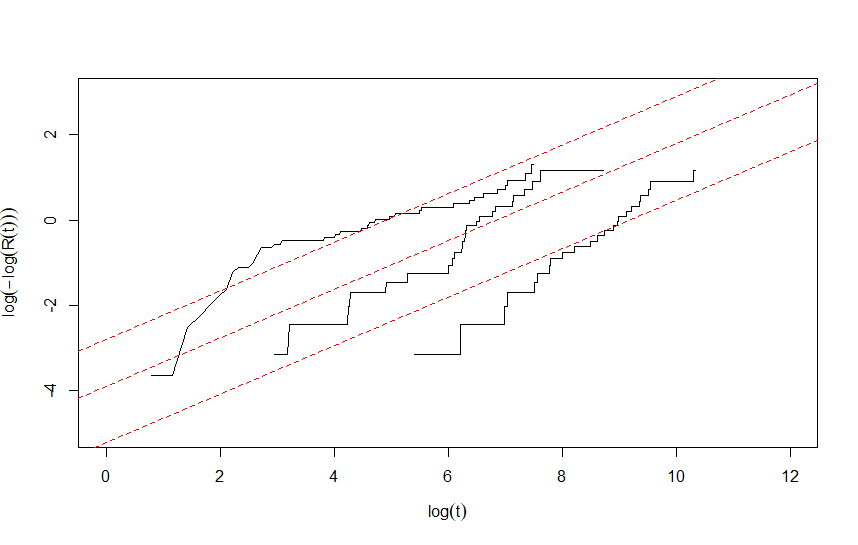}
\caption{Boxplots and a Weibull plot of the failure times of the pressure vessels under stresses of 25.5, 27.6 and 29.7 megapascals.}
\label{megabox}
\end{figure}

We consider an example from \cite{Ham08}. Interest lies in a reliability demonstration test plan for the time to failure of pressure vessels wrapped in Kevlar-49 fibres under stresses of 25.5 megapascals, 27.6 megapascals and 29.7 megapascals. Data have been collected in the form of failure times in hours of 87 such vessels over 8 different locations, known as spools. The data are given in \cite{Ham08} and boxplots of the natural logarithm of the failure times by stress level are given in the left hand side of Figure \ref{megabox}.


We see that there are quite strong differences between the distributions of time to failure under the different pressures. The highest stress has the shortest time to failure on average but also has the largest spread in failure times. Very few items fail quickly at the lowest pressure. The distribution of log times to failure at each of the pressure levels are fairly symmetrical. The median failure times in the three groups are 5194 hours, 543 hours and 55 hours respectively.

On the right hand side of Figure \ref{megabox} is a plot of $\log(t)$ against $\log\{-\log[\hat{R}(t)]\}$ for the failure times under each pressure individually, where $\hat{R}(t)$ is the empirical reliability at time $t$. If the Weibull distribution is suitable for the data, this should approximately represent a straight line. We see that a Weibull distribution appears to be suitable at each pressure. Also provided in the plot is a regression line plotted for the posterior means of the intercept and slope for each pressure. The gradients of the lines, which are estimates of $\beta$, were chosen to be the same in the model, but the intercepts, which are related to $\log(\rho)$, were allowed to be different for each pressure, and they are quite different. The lines appear to fit the data reasonably well for the three pressures. This provides justification for the choice of model structure.

We suppose that we are interested in the time beyond which 50\% of items survive, that is, the reliable life at $q=0.5$, and that we have a target for this of $\tau_{0.5,*}=4000$ hours at a stress of $s_{*}=25$ megapascals. In the design prior, we suppose that we have asked engineering experts for the quantities described in Section \ref{sec:priors} and they have given information which results in the specification of the prior hyper-parameters as $(\mu_0=-40, \mu_1=1, \sigma^2_0=1, \sigma^2_1=0.01, \sigma^2_{01}=0)$, $(a_\beta=20,b_\beta=13)$ and $(a_\epsilon=2,b_\epsilon=2)$. 

\begin{figure}[h!]
\centering
\includegraphics[height=3in]{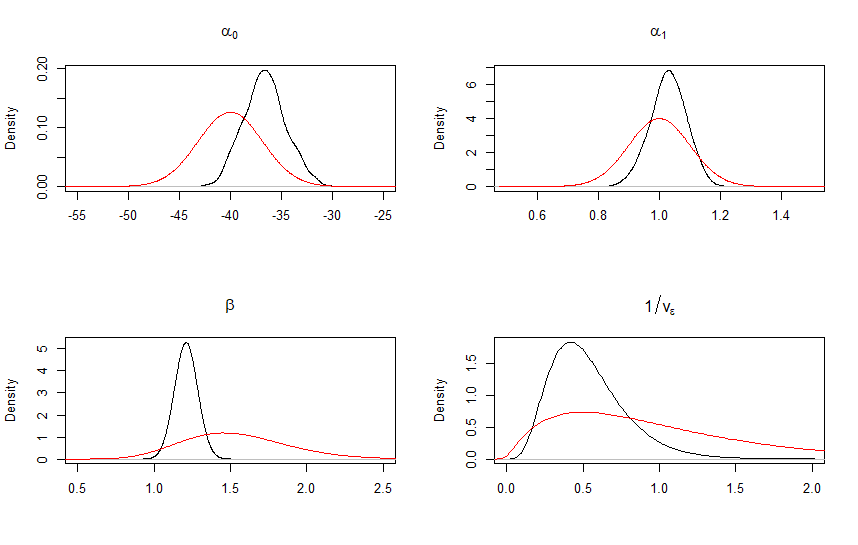}
\caption{The design posterior distributions (black) and prior distributions (red) of $\alpha_0, \alpha_1, \beta$ and $1/v_{\epsilon}$.}
\label{wei1}
\end{figure}

We can evaluate the design posterior distributions of the prior parameters $(\alpha_0, \alpha_1, \beta)$ and the prior precision of $\epsilon$, $1/v_{\epsilon}$, using MCMC. They are given in black, together with their design prior distributions in red, in Figure \ref{wei1}.
We see that the posterior distributions of each of the parameters are unimodal and relatively smooth. Each is more concentrated than their design prior distribution.

\begin{figure}[h!]
\centering
\includegraphics[height=3in]{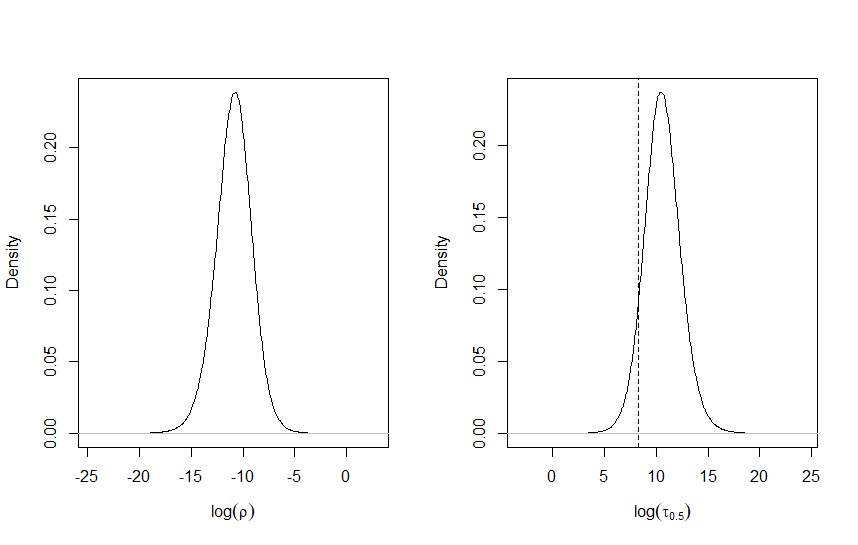}
\caption{The design posterior distribution of $\log(\rho)$ (left) and the log reliable life at $q=0.5$ (right).}
\label{wei2}
\end{figure}

Using these simulated values, we find the design posterior distribution of $\rho$ and the reliable life at $q=0.5$. Both are provided on the log-scale in Figure \ref{wei2} for a new location. The dashed line is the target, $\log(\tau_{q,*})=\log(4000)$.
We see that each posterior distribution is relatively symmetrical on the log-scale. The majority of the density for the log reliable life is above the target value.

We can use the simulations from the design posterior distributions for $\beta$ and $\rho$ to find the assurance for any choice of sample size $n$. To do so, we also need to define the analysis prior. In this example, we consider a sceptical analysis prior. The prior parameters are chosen to give $\Pr_A(\tau_{0.5}\geq4000)=0.1$. We suppose that the test is to be conducted at two accelerated pressures of 27 and 29 megapascals and that half of the vessels will be tested under each pressure. 

\begin{figure}[ht]
\centering
\includegraphics[height=3in]{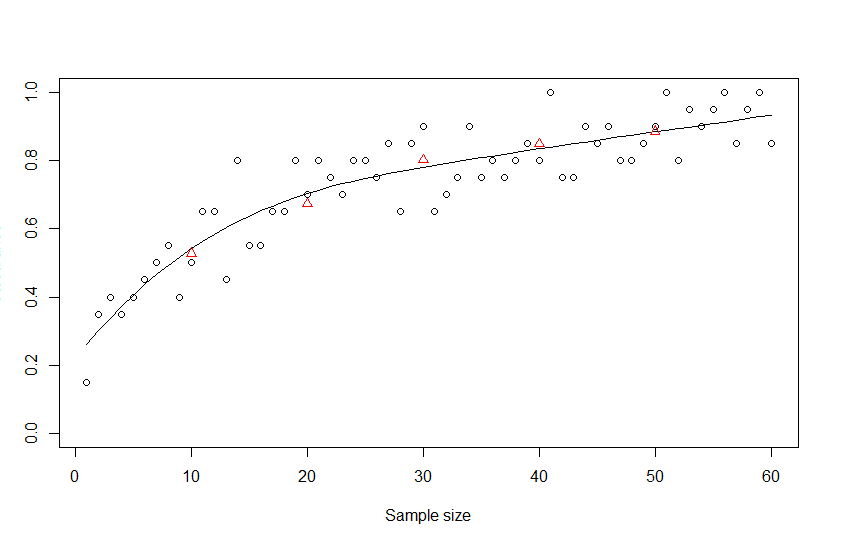}
\caption{The assurance based on the design posterior distribution for sample sizes in the range [1,60].}
\label{wei3}
\end{figure}

Using this analysis prior and test pressures, the assurance for various sample sizes, under the design posterior distribution, is given in Figure \ref{wei3}. To produce the assurance curve, 60 values of $n$ were chosen and the test criterion was evaluated 20 times for each value of $n$. The circular points on the plot are the empirical proportions for each chosen value of $n$. The estimated assurance was fitted using shape constrained additive models via monotonically increasing P-splines. The triangles on the plot are values of the assurance calculated from a very large number of Monte Carlo simulations.

We can achieve an assurance of more than 85\% with sample sizes under 60 using the design posterior distribution. The probability under the design posterior distribution that we meet the reliability target is around 89\%. Suppose that we wish to achieve an assurance of 80\%. Using the fitted curve, we can see that the first sample size to achieve this is $n=32$. So we would put 32 vessels on test, 16 under a pressure of 27 megapascals and 16 under a pressure of 29 megapascals. We would record each of the failure times, and, if, given those failure times, $\Pr_{A}(\tau_{0.5}>4000)\geq0.95$, then the test is passed. If not, the test is failed.

\begin{figure}[ht]
\centering
\includegraphics[height=3in]{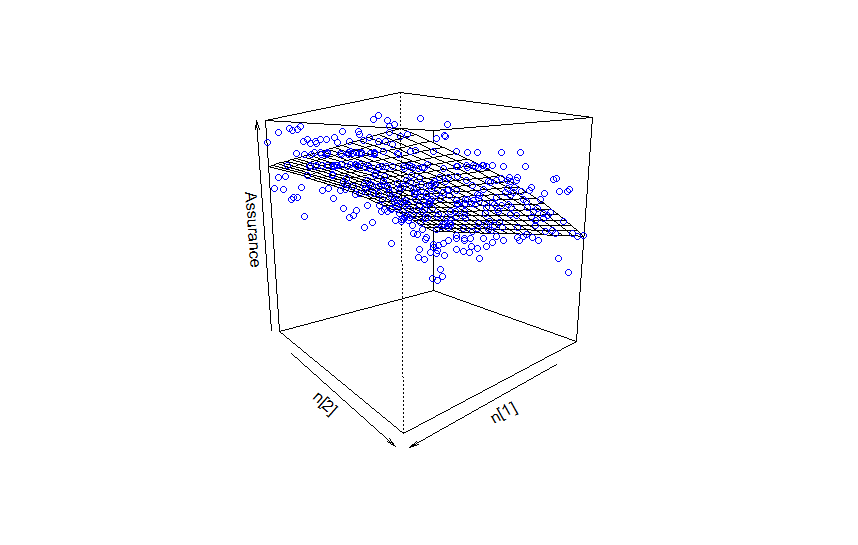}
\caption{The assurance based on the design posterior distribution for different combinations of $(n_{(27)},n_{(29)})$.}
\label{persp}
\end{figure}

Of course, testing half the vessels under each pressure may not lead to the optimal design. If we consider the pair of sample sizes $(n_{(27)},n_{(29)})$ for the number of vessels to test at $27$ and 29 megapascals respectively, we can evaluate the assurance using the curve fitting technique in 2-dimensions for all $n_{(27)},n_{(29)}\in 1,\ldots,20$, and the result, together with the simulated empirical proportions, is given as the surface in Figure \ref{persp}.

We provide the combinations of $(n_{(27)},n_{(29)})$ which give an assurance of at least 80\% with the smallest combined sample size and their estimated assurance in Table \ref{tab:ass}.

\begin{table}[ht]
\centering
\begin{tabular}{|cc|cc|} \hline
$n_{(27)}$ & $n_{(29)}$ & Total & Assurance \\ \hline
20 & 2 & 22 & 0.802 \\
20 & 3 & 23 & 0.808 \\
20 & 4 & 24 & 0.813 \\
19 & 5 & 24 & 0.804 \\
20 & 5 & 25 & 0.819 \\
19 & 6 & 25 & 0.810 \\
18 & 7 & 25 & 0.800 \\
20 & 6 & 26 & 0.825 \\
19 & 7 & 26 & 0.816 \\
18 & 8 & 26 & 0.806 \\ \hline 
\end{tabular}
\caption{Each RDT design giving an assurance of greater than 80\% sorted by total sample size.}
\label{tab:ass}
\end{table}

We see that we can achieve an assurance of greater than 80\% by putting just 22 items on test, 20 at 27 megapascals and 2 at 29 megapascals. This is fewer than the 32 observations when using equal sample sizes between the two pressures. However, we may wish to put 24 or 25 items on test to further increase the assurance by 1.1\% and 1.7\% respectively.

\section{Summary and further work}
\label{sec:conc}

In this paper we have considered the problem of sample size determination in reliability demonstration tests for hardware products leading to failure on demand and time to failure data. The approach we have taken is to use the concept of assurance, which chooses the sample size to answer the question ``what is the probability that the RDT will lead to a successful outcome?''. It can be used in conjunction with either frequentist or Bayesian analyses of the data following the test and, unlike typical risk criteria based approaches, separates the specification of prior beliefs in the design of the test from the prior to be used in the analysis following the test. Historical data can be incorporated into the sample size calculation without influencing the analysis of the RDT.

The methods have been fully developed in this paper, including advice on how to specify both the design and analysis prior in each case. However, in order for this approach to be adopted in practice, it will require extra resources to be available to those who plan and deliver RDTs. An important next step in this work will be to develop such resources. In particular, priorities are to incorporate the calculations in free open source software and to put together step by step guides to conduct the elicitations required to specify the design prior.  

The inference in the Weibull case has been performed using a combination of MCMC and a numerical scheme incorporating curve fitting. An adaption worth investigation is to develop an augmented MCMC scheme to perform the inference. We used a Bayesian analysis with a sceptical analysis prior for the time to failure application. Other options would be to perform the analysis based on a frequentist hypothesis test or to use a mixture analysis prior as in the failure on demand application.

The posterior risk criteria approach to RDT design is based on a single, shared prior distribution between the producer and the consumer. The approach could be adapted to incorporate separate priors for the design and analysis of the RDT. 

\bibliographystyle{plainnat}
\bibliography{refs}

\section*{Supplementary Material}

This development follows closely that in \cite{Ham08}. We define two different levels of reliability which indicate whether we are meeting reliability targets or not: $\pi_0$, the acceptable reliability level and $\pi_1$, the rejectable reliability level. If $\pi>\pi_0$ then the reliability is acceptable, if $\pi<\pi_1$ then it is not acceptable and if $\pi_1\leq\pi\leq\pi_0$ then we are in the indifference region. Most reliability demonstration testing has been based on these two reliability levels and making choices based on risk criteria. While classical risk criteria and average risk criteria have been proposed, we focus on Bayesian risk criteria. 

There are two errors we can make when making inferences based on the result of a reliability demonstration test. It could be the case that $\pi\geq\pi_0$ when a test is failed or that $\pi\leq\pi_1$ when a test is passed. We call the probability of these events the {\em producer's risk} and {\em consumer's risk} respectively. As part of the assurance testing we specify a maximum producer's risk $\alpha$ and a maximum consumer's risk $\beta$.

The posterior producer's risk can be calculated as follows.

\begin{eqnarray*}
\Pr(\pi\geq\pi_0\mid\textrm{Test failed}) & = & \Pr(\pi\geq\pi_0\mid y>c) \\
& = & \int_{\pi_0}^{1}p(\pi\mid y>c)d\pi \\
& = & \int_{\pi_0}^{1}\dfrac{\Pr(Y>c\mid\pi)p(\pi)}{\int_{0}^{1}\Pr(y>c\mid\pi)p(\pi)d\pi}d\pi \\
& = & \dfrac{\int_{\pi_0}^{1}\left[\sum_{y=c+1}^{n}{n \choose y}(1-\pi)^y\pi^{n-y}\right]p(\pi)d\pi}{\int_{0}^{1}\left[\sum_{y=c+1}^{n}{n \choose y}(1-\pi)^y\pi^{n-y}\right]p(\pi)d\pi} \\
& = & \dfrac{\int_{\pi_0}^{1}\left[1-\sum_{y=0}^{c}{n \choose y}(1-\pi)^y\pi^{n-y}\right]p(\pi)d\pi}{1-\int_{0}^{1}\left[\sum_{y=0}^{c}{n \choose y}(1-\pi)^y\pi^{n-y}\right]p(\pi)d\pi} \\
\end{eqnarray*}

Similarly, the posterior consumer's risk is given by

\begin{eqnarray*}
\Pr(\pi\leq\pi_1\mid\textrm{Test passed}) & = & \Pr(\pi\leq\pi_1\mid y\leq c) \\
& = & \dfrac{\int_{0}^{\pi_1}\left[\sum_{y=0}^{c}{n \choose y}(1-\pi)^y\pi^{n-y}\right]p(\pi)d\pi}{\int_{0}^{1}\left[\sum_{y=0}^{c}{n \choose y}(1-\pi)^y\pi^{n-y}\right]p(\pi)d\pi} \\
\end{eqnarray*}

Let us now suppose that we have data from previous tests of the form $x_i$ for $i=1,\ldots,I$ where $X_i\mid\pi_i\sim\textrm{bin}(n_i,\pi_i)$, where the probabilities of items surviving the test in each can be thought of as coming from the same prior distribution $\pi_i\sim\textrm{beta}(a,b)$, with hyperparameters $a,b$. We give a hyper-prior distribution to $(a,b)$. Then, let us suppose that our probability of interest also comes from the same prior distribution $\pi\sim\textrm{beta}(a,b)$.

We can then specify the producer's risk and the consumer's risk for the test plan $(n,c)$ as, respectively,

\begin{eqnarray*}
\Pr(\pi\geq\pi_0\mid\textrm{Test failed},\bm x) & = & \dfrac{\int_{\pi_0}^{1}\left[1-\sum_{y=0}^{c}{n \choose y}(1-\pi)^y\pi^{n-y}\right]p(\pi\mid\bm x)d\pi}{1-\int_{0}^{1}\left[\sum_{y=0}^{c}{n \choose y}(1-\pi)^y\pi^{n-y}\right]p(\pi\mid\bm x)d\pi} \\
\end{eqnarray*}
and
\begin{eqnarray*}
\Pr(\pi\leq\pi_1\mid\textrm{Test passed},\bm x)
& = & \dfrac{\int_{0}^{\pi_1}\left[\sum_{y=0}^{c}{n \choose y}(1-\pi)^y\pi^{n-y}\right]p(\pi\mid\bm x)d\pi}{\int_{0}^{1}\left[\sum_{y=0}^{c}{n \choose y}(1-\pi)^y\pi^{n-y}\right]p(\pi\mid\bm x)d\pi}. \\
\end{eqnarray*}

We can sample from the posterior distribution $p(\pi\mid\bm x)$ using Markov Chain Monte Carlo (MCMC) in the following way.
\begin{enumerate}
 \item Generate $N$ posterior draws of $a,b$ of the form $a^{(j)},b^{(j)}$ for $j=1,\ldots,N$.
 \item For $j=1,\ldots,N$ draw $\pi^{(1)},\ldots,\pi^{(N)}$ as
 \begin{displaymath}
 \pi^{(j)}\sim\textrm{beta}(a^{(j)},b^{(j)}).
 \end{displaymath}
\end{enumerate}
Using the draws of $\pi$ from the posterior distribution we can evaluate the producer's risk as
\begin{eqnarray*}
\Pr(\pi\geq\pi_0\mid y>c,\bm x) & \approx & \dfrac{\dfrac{1}{N}\sum_{j=1}^{N}\left[1-\sum_{y=0}^{c}{n \choose y}(1-\pi^{(j)})^y(\pi^{(j)})^{n-y}\right]I(\pi^{(j)}\geq\pi_0)}{1-\dfrac{1}{N}\sum_{j=1}^{N}\left[\sum_{y=0}^{c}{n \choose y}(1-\pi^{(j)})^y(\pi^{(j)})^{n-y}\right]} \\
\end{eqnarray*}
and the consumer's risk as
\begin{eqnarray*}
\Pr(\pi\leq\pi_1\mid y\leq c,\bm x) & \approx & \dfrac{\sum_{j=1}^{N}\left[\sum_{y=0}^{c}{n \choose y}(1-\pi^{(j)})^y(\pi^{(j)})^{n-y}\right]I(\pi^{(j)}\leq\pi_1)}{\sum_{j=1}^{N}\left[\sum_{y=0}^{c}{n \choose y}(1-\pi^{(j)})^y(\pi^{(j)})^{n-y}\right]}. \\
\end{eqnarray*}

A test plan is then the pair $(n,c)$ for a given $(\alpha,\beta)$ such that
\begin{eqnarray*}
\Pr(\pi\geq\pi_0\mid y>c) & \leq & \alpha, \\
\Pr(\pi\leq\pi_1\mid y\leq c) & \leq & \beta. \\
\end{eqnarray*}
Clearly there is more than one possible choice of test plan in this case. Approaches have been suggested to provide a smallest pair $(n,c)$. For example, we could choose the smallest $n$ such that there exists a $c$ which satisfies the two inequalities.

In practice, it is unlikely that we would observe $y>c$ or $y\leq c$. It is more likely that we would observe the actual value of $y$. Therefore an alternative to the posterior risk criteria would be to say that, for a given $n$ a value of $c$ acceptable to the producer would be one where, for all $y\leq c$, $\Pr(\pi\geq\pi_0\mid y)\leq\alpha$ and a value of $c$ acceptable to the consumer would be one where, for all $y\leq c$, $\Pr(\pi\leq\pi_1\mid y)\leq\beta$. The pair $(n,c)$ could then be chosen to satisfy the two inequalities. 

\end{document}